# Detectors and cryostat design for the SuMIRe Prime Focus Spectrograph (PFS)

James E. Gunn [1], Michael Carr [1], Stephen A. Smee [*,2], Joe D. Orndorff [2], Robert H. Barkhouser [2], Murdock Hart [2], Charles L. Bennett [2], Jenny E. Greene [1], Timothy Heckman [2], Hiroshi Karoji [4], Olivier LeFevre [3], Hung-Hsu Ling [6], Laurent Martin [3], Brice Ménard [2,4], Hitoshi Murayama [4], Eric Prieto [3], David Spergel [1], Michael A. Strauss [1], Hajime Sugai [4], Akitoshi Ueda [5], Shiang-Yu Wang [6], Rosemary Wyse [2], Nadia Zakamska [2]

[1] Department of Astrophysical Sciences, Princeton University, Princeton, NJ, 08544, (USA);
[2] Department of Physics and Astronomy, Johns Hopkins University, Baltimore, MD, 21218, (USA);
[3] LAM – Laboratoire d'Astrophysique de Marseille, Traverse du siphon, 13376 Marseille, (France);
[4] Kavli Institute for the Physics and Mathematics of the Universe, The University of Tokyo, (Japan);
[5] National Astronomical Observatory of Japan, Mitaka, Tokyo 181-8588, (Japan);
[6] Academia Sinica Institute of Astronomy and Astrophysics, Taipei 10617, (Taiwan)

## ABSTRACT

We describe the conceptual design of the camera cryostats, detectors, and detector readout electronics for the SuMIRe Prime Focus Spectrograph (PFS) being developed for the Subaru telescope. The SuMIRe PFS will consist of four identical spectrographs, each receiving 600 fibers from a 2400 fiber robotic positioner at the prime focus. Each spectrograph will have three channels covering wavelength ranges 3800 Å – 6700 Å, 6500 Å – 10000 Å, and 9700 Å – 13000 Å, with the dispersed light being imaged in each channel by a f/1.10 vacuum Schmidt camera. In the blue and red channels a pair of Hamamatsu 2K x 4K edge-buttable CCDs with 15 um pixels are used to form a 4K x 4K array. For the IR channel, the new Teledyne 4K x 4K, 15 um pixel, mercury-cadmium-telluride sensor with substrate removed for short-wavelength response and a 1.7 um cutoff will be used. Identical detector geometry and a nearly identical optical design allow for a common cryostat design with the only notable difference being the need for a cold radiation shield in the IR camera to mitigate thermal background. This paper describes the details of the cryostat design and cooling scheme, relevant thermal considerations and analysis, and discusses the detectors and detector readout electronics.

## 1. INTRODUCTION

The Sumire PFS will consist of 4 identical spectrographs each receiving 600 fibers from the 2400 fiber prime focus Cobra placement robot. Each spectrograph will have three channels, covering the wavelength ranges 3800-6700 Å, 6500-10000 Å, and 9700-13000 Å. The cameras for each channel are almost identical vacuum Schmidts with a focal ratio of f/1.10. We have constructed cameras of this general design for the double spectrograph on the 5-meter at Palomar [1] and for the echelle on the ARC 3.5-meter [2] with success.

Since we are not going very far into the infrared, we can (with care) use a room-temperature spectrograph, but must be very careful of the in-beam thermal background and design the camera/dewar for the IR detector so as to minimize the diffuse background. To go farther into the IR we would need to use a multi-spectrograph design with a cooled IR instrument such as the ESO X-shooter [3], but it is in any case not possible with realistic optics to reach the background levels we require in the presence of the very bright OH lines in the H band. The detectors for the optical channels will be Hamamatsu 2K x 4K fully-depleted CCDs, 200 um thick, with 15 um pixels, running at -110 C (163 K). These devices have been described in Kamata *et al*. (2006) [4] and their performance evaluated in Kamata *et al*. (2010).[5] Two of these will be side-butted in each camera to yield a 4K x 4K detector.

---

[*] smee@pha.jhu.edu; phone 410-516-7097; fax 410-516-6664

The infrared channel will have a 1.75-micron cutoff Teledyne H4RG-15 Mercury-Cadmium-Telluride device which is also a 4K x 4K device with 15 um pixels, physically the same size as the CCD pair. This detector will run at 110 K, much colder than the CCDs. The device is described in a recent paper by Blank *et al* (2011).[6]

It is the responsibility of Princeton University and Johns Hopkins University to build the systems, both mechanical and electronic, to mount and operate these detectors, and to build the cryostats which house the detectors and camera optics. This section deals with this subsystem. Since the camera optics is housed in the cryostat and the detectors interface directly with the camera optics and must do so to very high precision, the interface with camera optics provided by the Laboratoire d'Astrophysique de Marseille (LAM) is complex. We will describe here a fairly clean way to deal with this, though there are doubtless details which are not fully understood yet.

The optical/cryostat design is somewhat unusual, though there are a small number of systems in use which use the same principles. We need a very fast beam, since the scale on the sky is determined entirely by the camera f/ratio. Subaru is a very big telescope (8.2 m), and to use 15 um pixel devices efficiently (with reasonable sampling) one must use a *very* fast camera. One arcsecond at f/1.0 projects to 40 microns on the focal plane. We will use an f/1.1 camera and fibers which are about 1.1 arcseconds in diameter, and use fibers which have approximately 10% FRD loss, and all of this together makes a spot about 54 microns on the device, about 3.5 pixels. It is not possible with reasonable designs to make the currently fashionable refracting systems this fast, and modified Schmidts are the only reasonable choice. In a fast Schmidt, the detector must be in the beam, and so obscures light, so it is desirable to have the detector package as small as possible, preferably no larger than the detector itself. This is most easily accomplished by making the cryostat the camera housing, so the cold detector and all the reflective optics are in vacuum. The Schmidt design lends itself to this, since there is a corrector plate which can be made thick which can serve as a vacuum window. This design also can accommodate a cold thermal shield, which turns out to be necessary for good performance with the IR detector.

The subsystem has many components, which will be discussed below. There are 12 complete subsystems for the 4 spectrographs. The systems are designed to be as nearly identical as possible, given the rather different thermal requirements for the CCDs and IR detectors. The components include:

- **The cryostat body:** This is the mechanical assembly which is the envelope for the whole system for one camera. The LAM camera assembly forms a rigid optical assembly, so the mechanical requirements on the envelope are not especially severe, but it must be vacuum-tight, strong, provide the proper thermal environment for the detectors, and provide connectors for the detector and thermal electronics and any necessary mechanical access to the dewar interior.

- **The detector mountings:** This includes the mechanical interface to the camera optics supplied by LAM, quite possibly a tilt/focus mechanism, either manual or electric, and the fixturing required to install and remove the detectors safely and accurately. The details of this will be different for the CCDs and the IR detectors.

- **The detectors themselves:** These are purchased items, but are a major component of the cost; their performance to a very large extent determines the performance of the spectrograph. The care with which they are mounted and run is a very large factor in determining this performance.

- **The thermal system:** This includes the mechanical refrigerators to cool the detectors, various heat-strapping components to provide conductive paths from the detectors to the coolers, and a rather complex thermal shielding system which does the double duty of providing a cold radiation environment for the detectors, particularly the IR ones, and reducing the thermal load so as to be able to provide the necessary cooling without unreasonable cooler requirements. Also part of this system is the temperature-regulating electronics to maintain the detector temperatures to the necessary precision.

- **The vacuum system:** This includes the valving for the cryostat, pumping facilities for creating and maintaining the vacuum, and facilities for monitoring the vacuum.

- **The detector electronics system:** This includes the electronic systems in the dewar which directly interface to the detectors and the electronics outside which receive the signals from the detectors/in-vacuum electronics and provide an interface to the Subaru data system.

- **The utilities electronics system:** This subsystem, which is actually a set of smaller systems, some of which have already been named above, includes the electronics for temperature control, vacuum monitoring, focus, and any other "housekeeping" functions such as monitoring the wall temperatures, temperatures of the cooler bodies themselves and the power supplies, monitoring cooler power, perhaps cooler vibration, etc.

## 2. CRYOSTAT BODY

The camera assembly which must be housed in the cryostat is being designed and produced by LAM; the assembly is shown schematically in Figure 1. The Schmidt corrector is held in the large ring, called here the front bell, of the assembly. The dark grey ring is a separate part, accurately indexed to the bell and hence to the corrector, but removable for insertion of the detector, wiring, and heat shield, all of which will be discussed later. The primary mirror is supported off this ring by the hexapod shown, invar tubes with a G10 section and titanium flexures. The thermal isolation to the front bell, which is, of course, at room temperature, is provided by this hexapod. The primary mirror will be made of quartz or Zerodur; the tripod which supports the detector assembly, and the detector assembly housing (the open rectangle at the apex of the tripod) are fabricated of invar. The tripod legs will have jackets of OFHC copper, since the heat from the detector will be conducted along this path. LAM will fabricate and align this assembly and inform PU/JHU of exactly where the focal plane is; it is our responsibility to place the detector front surface in that plane.

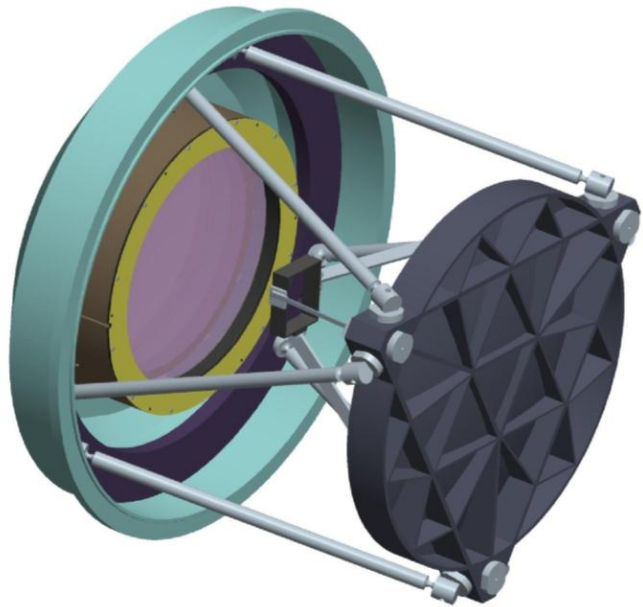

Figure 1  The camera assembly which will be delivered by LAM.  The cryostat must enclose this assembly and provide a proper vacuum and thermal environment for the detector, which is mounted in the box at the top of the tripod.  Not shown is the field flattener lens, which is mounted on the front (the side facing the primary mirror) of the box.

The assembled cryostat with the camera installed is shown in cross-section in Figure 2. The details of the front bell are still somewhat uncertain, since neither the optical design nor the design of the spectrograph body is yet final, so what is shown is very schematic. Here one sees the interface ring from which the hexapod is suspended, and the outside "can" of the dewar. This can consists of three parts: A front ring, which carries the hermetic feedthrough connectors for all the electronics and the focus mechanism. Behind this ring is the rear tube, from which there are no electrical connections to anything in the camera assembly. It does have two thermal connections, one to the mirror/tripod assembly on the back plate, and another to the thermal shield on the side. Access ports allow disconnection of these, so that it can be removed without disturbing the camera assembly at all. The vacuum space is closed off at the rear by a plate which carries the cryocoolers, the vacuum pumping ports, and the ion pump. The only connection of this plate to the camera assembly is the aforementioned thermal path to the cryocooler, which we discuss later. The assembly is designed to allow easy assembly and disassembly, and replacement of a failed cooler without dismounting the camera. The material for the cryostat is aluminum; the total mass of the cryostat assembly complete is 200 kilograms.

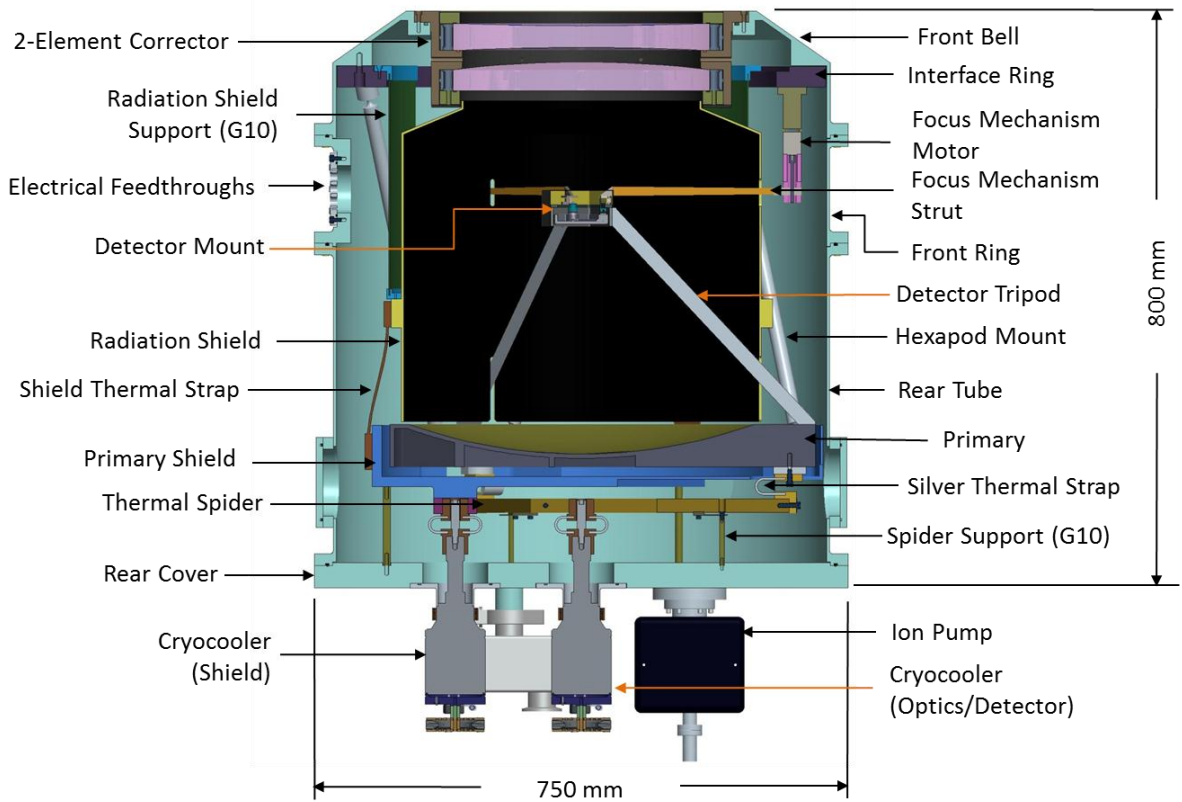

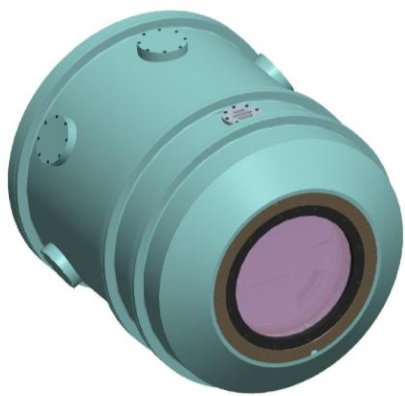
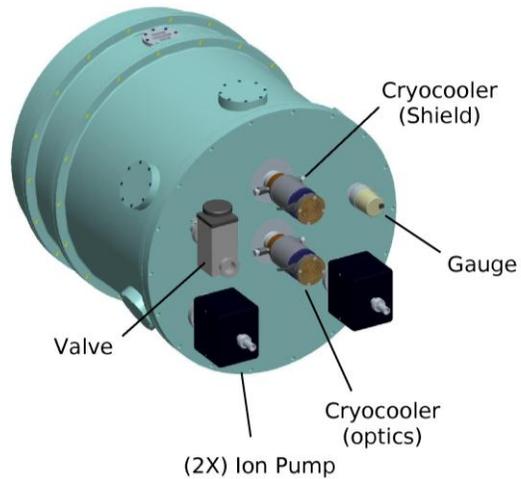

Figure 2 The assembled cryostat. The top panel is a cutaway section with the major components labeled and a scale for size. The bottom shows the assembly from the front and rear, showing the coolers, valves, and ion pump.

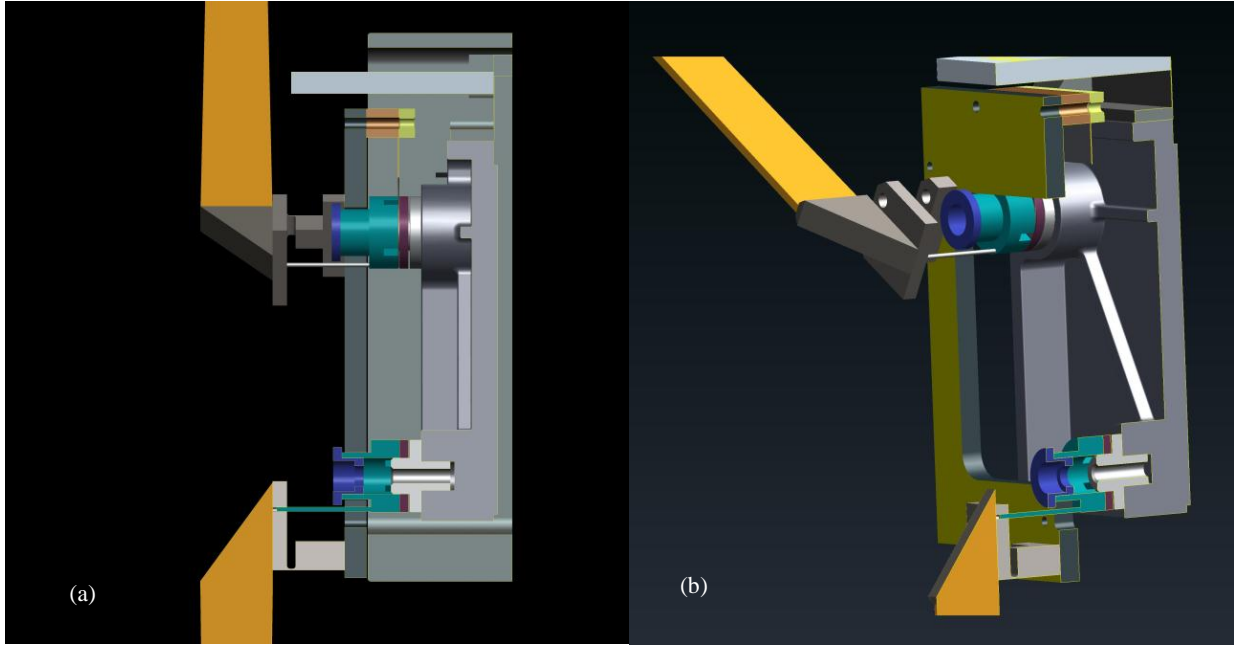

Figure 3 (a) Cross-section view of the IR detector mounting. The detector is located by three precisely located referece posts. Two of these are picked up by the thin flexure shown in the diagram. The green pieces clamp the flexure to the detector mountings and provide the detector-end anchor for the flexural rods coming from the focus adjustment levers. These rods provide fine tip-tilt and focus adjustments; the flexure locates the detector transversely and in rotation. And (b) Cutaway of the same view. One can see a little better the arrangement of the mounting. The screws which hold the green clamps to the detector mounting bosses and the others which hold the whole assembly together are not shown. The lever arms, shown schematically here, will be thin G10 plates.

The evacuation of the vessel produces very large forces on the walls. The back plate has an applied force of 2.4 metric tons; the deflection of that plate (if it is 37 mm thick 6061) is 0.2 mm when the vessel is evacuated. The force on the front bell is as large and opposing. This would move the corrector toward the primary by an amount comparable to the bow in the back plate, but this movement can be lessened by the design and attachment of the interface ring, and in any case can be calculated and allowed for. We or LAM will do a full FEA, but this awaits a bit more maturity of the optical and mechanical design. The design and construction of the cryostat body is quite straightforward and does not pose any significant risk. We plan to use hermetic high-density D-sub connectors for all connections through to the vacuum. These are available in satisfactory configurations with up to 78 pins; we will have to use no more than 2 of them on the dewar for the detector connections and perhaps another smaller one for utility wiring.

## 3. THE DETECTOR MOUNTINGS

The detector is mounted within the small invar box bonded to the top of the detector tripod. The camera assembly provides a flat mounting surface on the rear of these boxes and six tapped holes for fastening the invar detector mounting plate. This surface is the reference surface for the detector; the focal plane is a known distance in front of this surface. We will use an active focus mechanism with a travel of perhaps ± 100 microns. In the drawings shown here it is implemented with three Phytron VSS-25 vacuum-compatible stepping motors mounted on the front ring and directly driving screws which move very lightweight levers aligned with the tripod legs, though we will probably use ferrofluidic feedthrough couplings and move the motors out of the vacuum. With reduction of about 100:1, 10 mm of travel on the lever moves the detector 100 microns. With three levers like this, the piston and tilt of the detector can be adjusted for focus, evaluating optical performance, taking up any changes caused by changing environmental temperature, disassembly/reassembly, or any other less than perfectly understood environmental/setup conditions. The mountings for the IR detectors are shown in cross-section and exploded view in Figure 3a and Figure 3b, respectively. The CCD mountings are very similar, but are a little more complex since there are 2 coplanar detectors mounted on single plate which is moved by the focus mechanism. The Teledyne H4RG-15 is not square but has a protrusion on one side to handle the output connectors. This is unfortunate in that it costs light, but is not, apparently, possible to change. For these reasons the detector mountings are not identical, but the principles and many of the parts are. The detector is surrounded by (in front) the field flattening lens, (the sides) the invar box, and (in back) the invar mounting plate. It is suspended from a thin bronze flexure, seen most clearly in the cross-section, attached to one set of mounting points. This flexure constrains the detector in rotation and transverse motion, but leaves it free to execute small motions in tip and tilt. These are controlled by the 100:1 levers pivoting on the flexural hinges shown in the drawings and connected to the detector mounting points by thin metal rod flexures. Even though the detector is surrounded by cold surfaces, there is some dissipation in it, and it is connected thermally to the outer structure with silver strap flexures, which will keep it within about 2 degrees of the cold surrounding.

As we shall see in Section 5, we will use a composite invar/OFHC copper tripod supporting the detectors which have legs which are 5 mm thick and 20 mm wide. The invar supplies the stiffness and metering; the copper, which is not mechanically rigidly affixed to the invar legs, provides the thermal path from the detector to the thermal connections to the cooler. Given this geometry and the geometry of the detector mountings, we can calculate the obscuration this represents in the input beam. Clearly, as one goes off-axis, the obscuration increases, because one begins to look at both the detector box and the tripod legs obliquely instead of face-on. The camera has a full field of about 18 degrees, so this effect is not small. For the CCDs, the obscuring area on axis is 13% for the CCD dewars, slightly larger (because of the larger detector package) for the IR dewars, 15%. This rises to 15 and 17 percent, respectively, at full field.

## 4. THE DETECTORS

### 4.1 The CCDs

As mentioned in the introduction, the detectors in the visible (red and blue) arms will be Hamamatsu 2K x 4K 15 um pixel fully-depleted CCDs of the same type which are being used in HyperSuprimeCam, except that we will use devices with better coatings which have better blue sensitivity in the blue arm. We will use fully-depleted devices because they have good quantum efficiency well into the deep red, with QE of order 50 percent at operating temperature at 1 micron. QE curves for the HSC CCDs are shown in Figure 4, which also shows the blue-enhanced two-layer coating, though we are hoping to get a special single-layer coating optimized at 5000 Å which would provide even better QE in the blue arm, since we do not care about the red response; we are currently investigating this coating with Hamamatsu. We will also use thinner devices for the blue arm, as we discuss below.

A drawing of the device on its AlN substrate is shown in Figure 5. The devices were built to HSC specs, which are also satisfactory for PFS. These specifications and the measured performance for HSC are shown in Table 1. These performance specs are for a readout rate of 150 KHz; we will read the devices at 75 kHz. This will reduce the readout noise to the vicinity of 3 electrons, which is the number upon which the strawman performance in the White Paper was based. Tests at NAOJ confirm that the read noise is what is expected from an output transistor in the white noise regime, so that the noise goes up as the square root of the pixel read frequency. The measured linearity residual of the HSC devices is shown in Figure 6. The indicated very low-level nonlinearity is small, nowhere larger than half a percent, but is somewhat worrying given the necessary accuracy of our sky subtraction. Our signals will typically be in the range of a few tens to a hundred electrons per pixel, and we will certainly digitize the signal with a resolution of the order of 1 electron per ADU, so we will saturate the 16-bit A/D converters well below the physical full well.

Each CCD has 4 amplifiers, each addressing a 512 x 4096 (usable) pixel area on the CCD. Thus the system will have 8 outputs. At a readout rate of 75 kHz, the read time is then 28 seconds. We will run the spectra vertically on the chip, so that the spectra run along the columns of the CCD. The small gap between the two CCDS then runs in the wavelength direction and can be avoided by placing the fibers with a gap on the slit - and the gap does *not* cause a wavelength gap in *all* the spectra. The chips will be run between -100 C and -110 C. The measured dark current performance at NAOJ is shown in Figure 7. At -100 C, the dark current is about 2.5 electrons per pixel per hour, which, for the typical 20 minute exposures on PFS will result in less than one electron per pixel per exposure. With a read noise of 3.2 electrons, this is a total noise of 3.33 electrons, so the dark current contributes negligibly to the system noise. Since the red QE in the 1 micron region is a fairly sensitive function of temperature, we will run the red chips as warm as we can in order to keep the far red sensitivity high without incurring too much noise from dark current or being severely impacted by hot pixels;

this will be an experimentally determined setting. It may well be, as has been our experience with the LBNL BOSS CCDs, that the necessary temperature is set by the necessity to freeze out hot pixel defects, and we should be prepared to take the QE hit if that is the case.

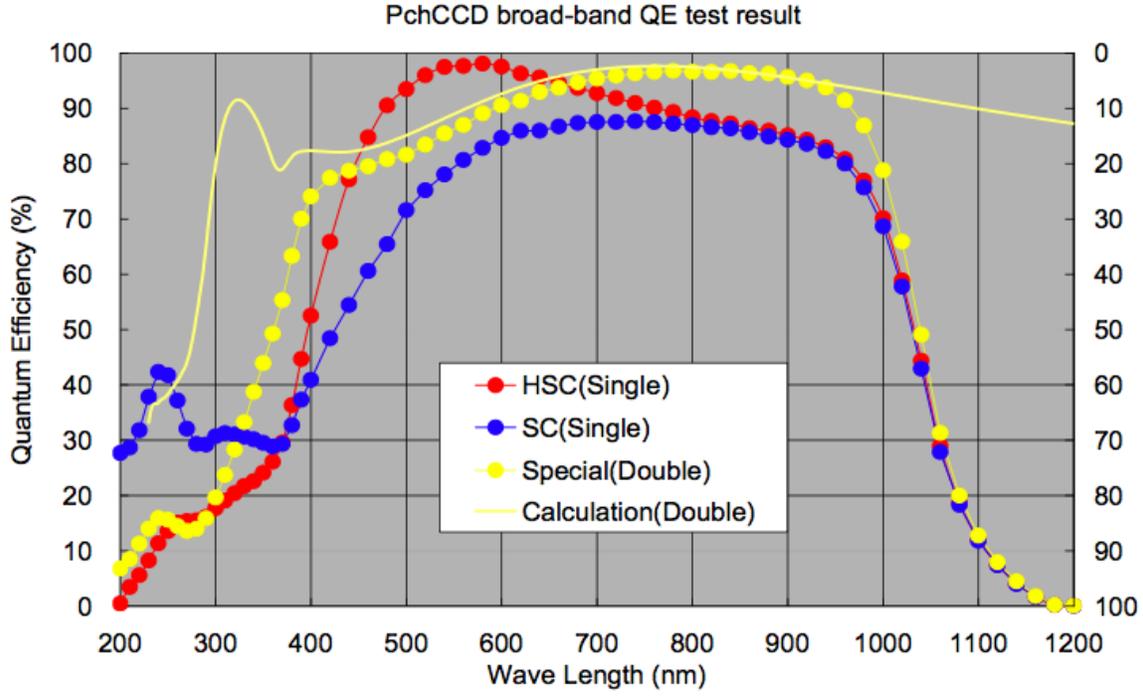

Figure 4  The quantum efficiency of the HSC CCDs, along with the earlier results for the less-efficiently coated SuprimeCam CCDs, and a developmental blue coating which we *may* be able to obtain for the chips for the blue camera.

Table 1  The quantum efficiency of the HSC CCDs, along with the earlier results for the less-efficiently coated SuprimeCam CCDs, and a developmental blue coating which we be able to obtain for the chips for the blue camera.

| Items | | Requirement (-100°C) | Measured |
|---|---|---|---|
| Packaging | Format (pixel size) | 2048×4096 (15 $\mu$m□) | - |
| | Pixel to Package edge | < 0.5 mm | 0.410±0.025 |
| | (Serial register side) | < 5.0 mm | 4.975±0.025 |
| | Global height variation | < 25 $\mu$m Peak-to-Valley | |
| QE | 400 nm | > 45 | 42 |
| | 550 nm | > 85 | 87 |
| | 650 nm | > 90 | 94 |
| | 770 nm | > 85 | 91 |
| | 920 nm | > 80 | 78 |
| | 1000 nm | > 40 | 40 |
| CTE (per pix) | Parallel direction | > 0.999995 (1600 e) | 0.999999 |
| | Serial direction | > 0.999995 (1600 e) | 0.999998 |
| Dark Current | | < a few e/hour/pix | 1.4 |
| Charge diffusion | | $\sigma_D$ < 7.5 $\mu$m (400 < $\lambda$ < 1050 nm) | 7.5 |
| Full well | 1 % departure | > 150,000 e | 180,000 |
| Amp. Responsivity | | > 4 $\mu$V/e | 4.5 |
| Readout noise | 150 kHz readout | < 5 e | 4.5 |

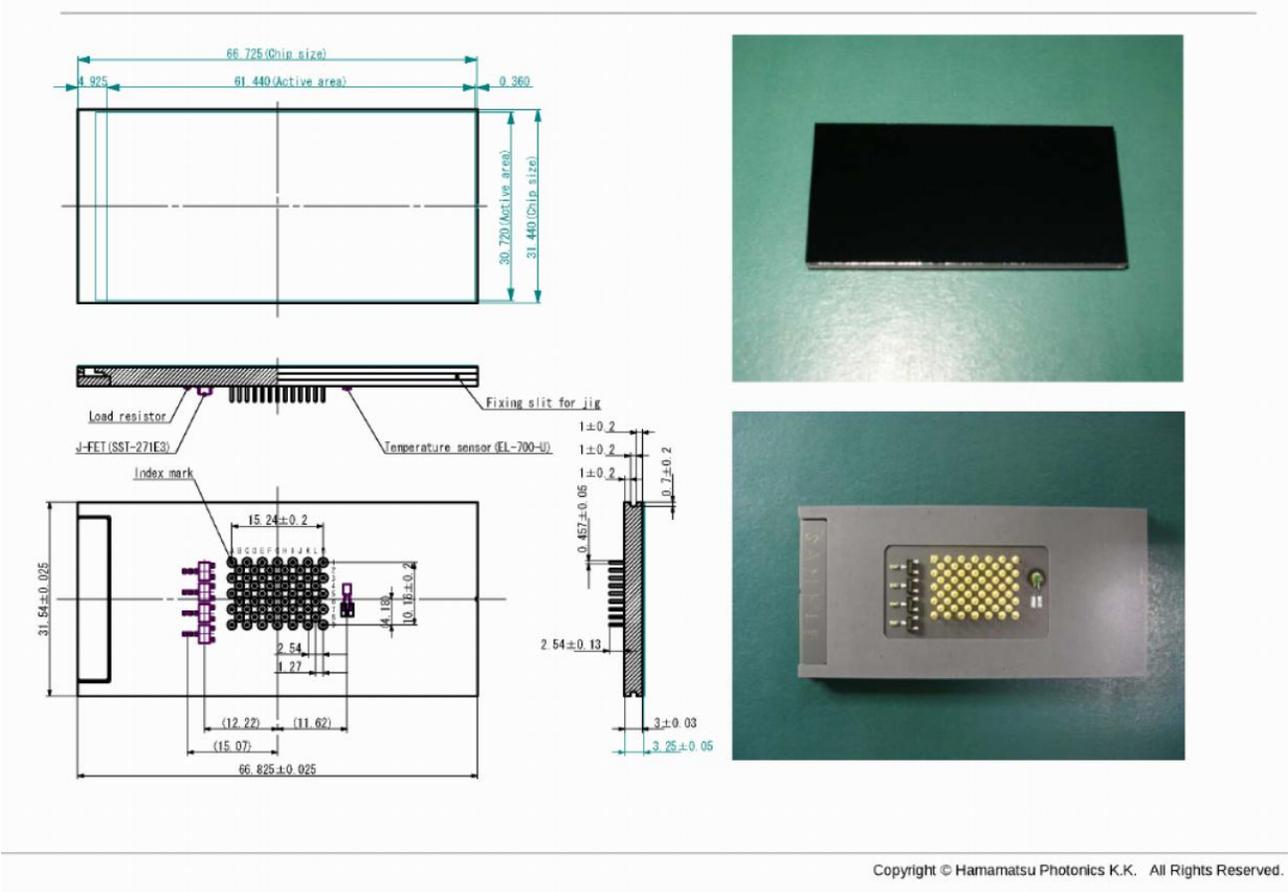

Figure 5  Dimensioned drawing of the 2Kx4K Hamamatsu HSC CCD, with two photos of the device.

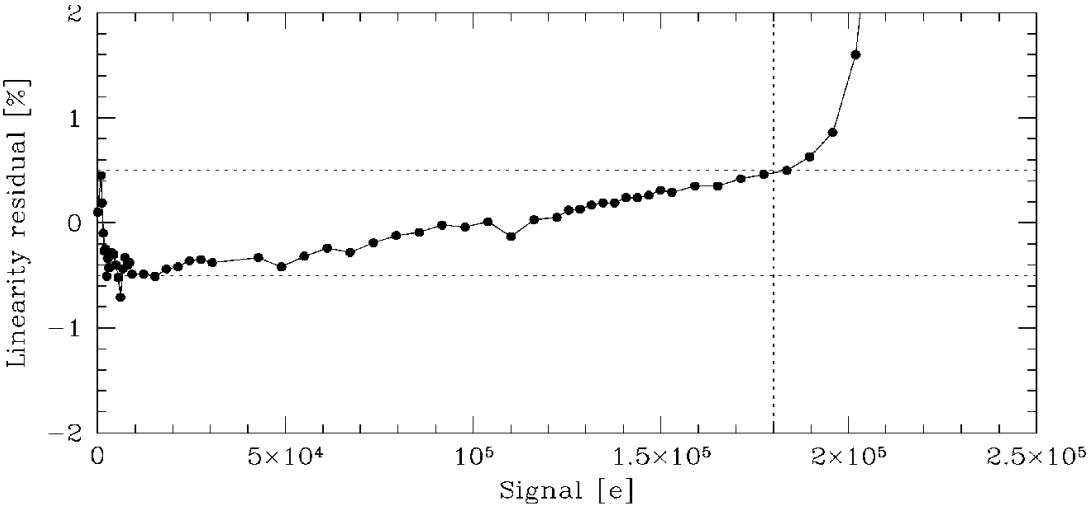

Figure 6  Measured linearity residual from the HSC team for the Hamamatsu CCDs.  We will digitize at a level such that the 16-bit A/D converters saturate at about 65-70,000 electrons, well below the onset of physical saturation.

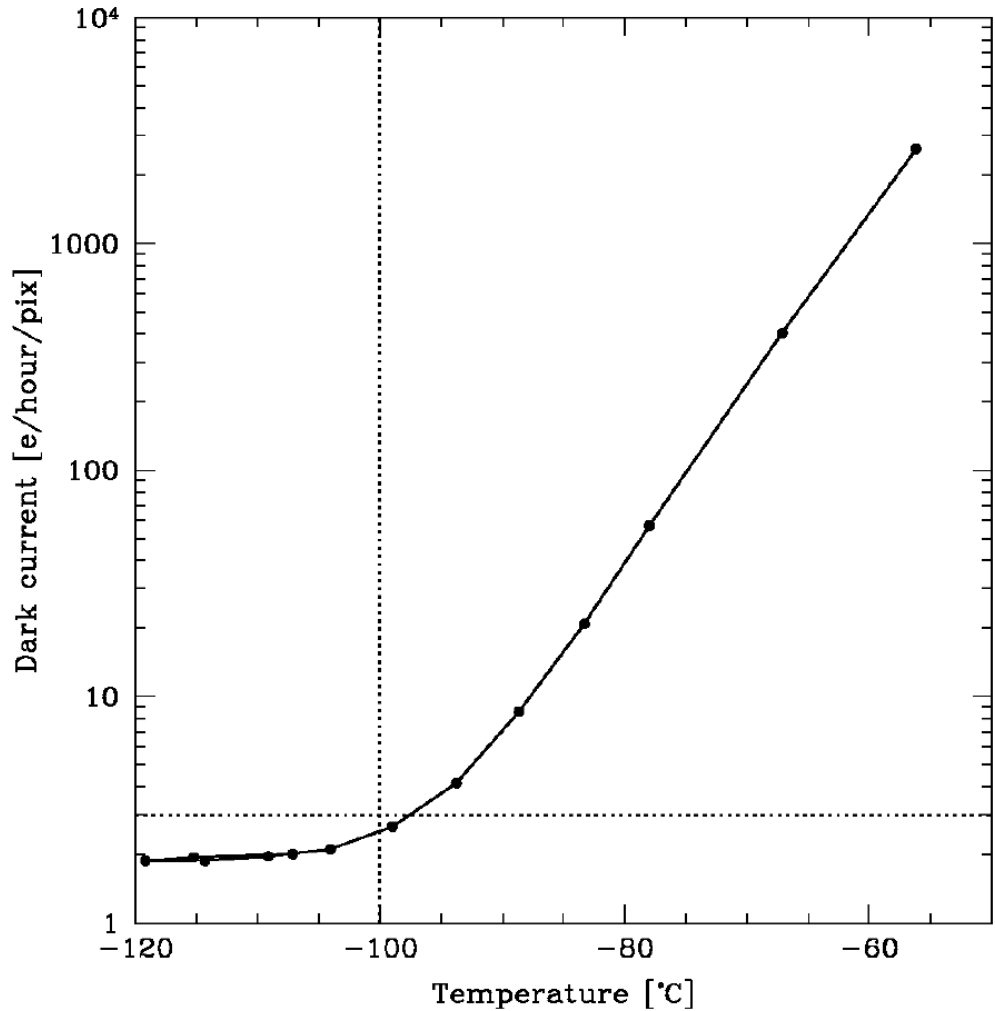

Figure 7  Average measured dark current of the Hamamatsu device. The real operating temperature will be determined by the necessity to freeze out defects, but certainly operating near -100 C is satisfactory for most pixels.

These are thick devices (200 microns for the HSC chips; this thickness is satisfactory for us as well in the red arm, but we will use 100 micron thick devices in the blue), which exhibit charge diffusion for photoelectrons (holes, actually; these are p-channel devices) which are liberated near the photosurface of the device. To minimize this, the devices are run with a large DC bias voltage to create an electric field in the device to accelerate the holes toward the collection electrodes. There is still finite diffusion, however, which in these devices results in a roughly gaussian charge spread with a $\sigma$ of about 7.5 microns, or a FWHM of 17.6 microns, a little larger than a pixel for the red chip, and about 12.5 microns, a little smaller, for the blue. Accounting for this requires some care. The rms diameter for a gaussian is 2.8 times $\sigma$, or 21 microns, a little larger than the FWHM. The fiber footprint, however, is a uniformly illuminated disk, 56 microns in diameter, which has an rms diameter 0.7 times its outer diameter, or 40 microns. It is the RMS diameters which add, approximately, so the degradation in resolution is close to 12 percent in the red, and about 5 percent in the blue. The degradation in the red is serious; we do not have a final optical design yet, and all of the designs we have looked at perform rather better in the red than in the blue, so it may not be a serious problem, but we must take it into account. We return to this issue after we discuss the effect of transparency below. For *red* photons, there is another effect. The chips become semi-transparent at the red end (and in fact that is why the QE drops off in the red shortward of the bandgap energy). Thus the photons on average travel some considerable distance into the device before they are absorbed; as a result, if the focus is placed on the chip surface, the beam has spread significantly by the time an average photon is absorbed. This effect is greatly mitigated by the fact that the refractive index of silicon is very large ~ 3.6) at the wavelengths of interest, so the very fast f/1.1 input beam becomes about f/4 in the CCD. The effect of this is readily calculated; one simply integrates the mean square geometric size of the beam at some height $h$ with the focus placed at a given height $h_0$ (either at the surface or within the detector) multiplied by the probability that the photon is absorbed at that height.

This is not the whole story, however, because the charge diffusion operates in the opposite direction; red photons are deposited deeper in the device and suffer less diffusion spread than blue ones, which are deposited very near the surface. Thus one has to add to the geometric size the diffusion size, which (in the mean square measure) is just the total diffusion size squared multiplied by the fraction of the chip thickness remaining at a given absorption height.

This total mean square size can then be minimized as a function of the focus position $h_0$. This best focal surface is quite strongly curved, but, in fact, a best-fitting plane does a quite good job. The situation is summarized in Figure 8. The diffusion dominates the image size at all wavelengths, and the variation with wavelength is not large. It is to be noted, however, that this ~ 20 micron degradation of the image diameter results in a roughly 12 percent degradation of the resolution of the spectrograph, as discussed above. It would be very good to reduce the diffusion losses, which one could do by increasing the substrate bias voltage. It is my understanding that Hamamatsu is evaluating how to enable this, and we should follow these developments carefully.

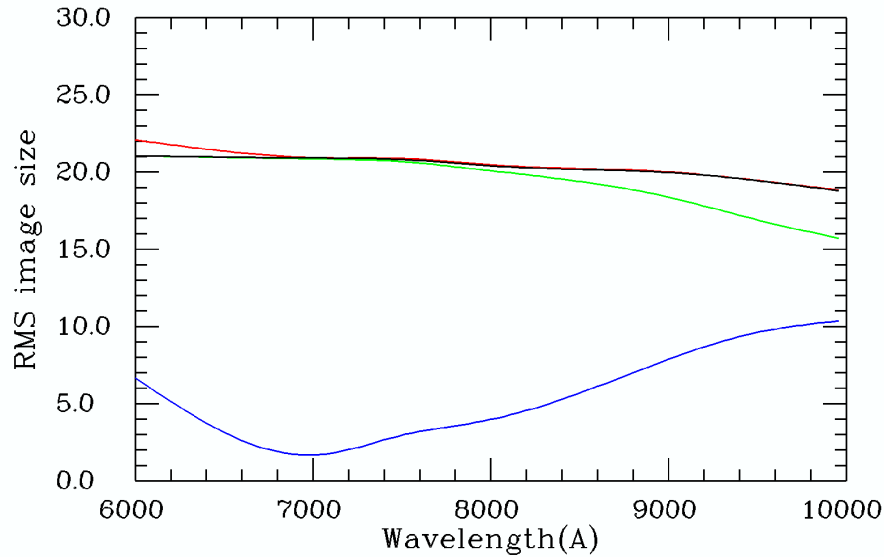

Figure 8 The effects of chip transparency and charge diffusion on image size for the CCD detectors. The contributions to the RMS image size (microns) are shown. The black line is the total RMS size for the best focus position on the chip, which is at the surface of the chip in the blue, increasing steeply to 25 microns (in air - the depth in the silicon is larger by a factor of 3.6) at 1 micron. This surface is strongly curved, but if one fits the best plane to it, the resulting RMS size is shown with the red line. The green and blue lines are the contribution from charge diffusion and the geometric effect of the transparency of the chip, respectively.

The flatness of the Hamamatsu detectors, as determined from the very large number of chips delivered to HSC, is not as good as one might desire. The typical peak-to-valley range in the flatness is about 20 microns, though a substantial fraction is better than this. This will be an area of concern for us. A full analysis of the HSC data to determine the statistics and nature of the departures from flatness will be done shortly, and Hamamatsu have shown much improved (about a factor of two better) flatness with recent process improvements.

## 4.2 The IR detectors

The IR detector chosen for the infrared arm is the Teledyne Hawaii 4RG-15 HgCdTe device made with 1.75 micron cutoff material; there is really no alternative available, and the device appears to be a really superb match to our requirements. The geometry of the active area of the device is essentially identical to the pair of Hamamatsu CCDs we will be using: an array of 4096 x 4096 15 um pixels. As already noted, the package is *not* square, but has an approximately 1 cm extension for connectors and bypass capacitors on one edge. This costs us a bit of light (about 2 percent). A photo of the device is shown in Figure 9, along with the ``Sidecar'' signal-processing ASIC chip we will discuss in the electronics section. The sidecar is the only electronics module in the vacuum necessary for these detectors.

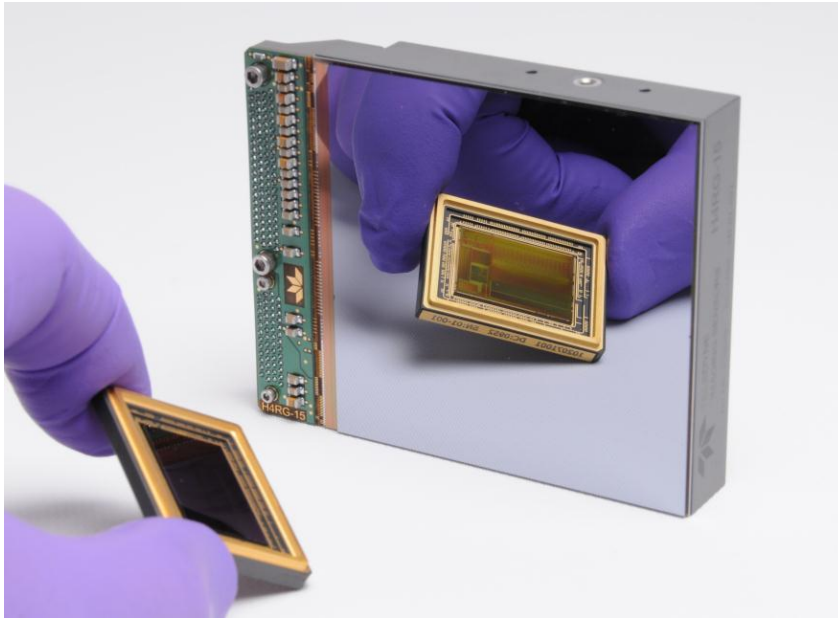

Figure 9 Photograph of the Teledyne H4RG-15 showing the circuitry on one edge. The gloved hand in the foreground is showing the new hermetic package for the sidecar signal processing ASIC, which we discuss below in the text.

The H4RG has the following characteristics which are of interest to us:

- **Quantum efficiency:** The measured QE is approximately 87 percent, very constant, over the wavelength range of interest for us. A figure showing QE for 4 tested devices is given in Figure 10. Of some interest to us is the quantum efficiency at longer wavelengths than the cutoff because of the very high thermal background photon flux. This may be an item of some concern, but measurements made at ESO with similar devices show no measureable response beyond the cutoff. This appears to be a desirable property of devices made with molecular beam epitaxy (MBE) as these devices are. We will see below that even so, we will need very strong filtration between our cutoff at 1300 nm and the chip cutoff at 1750 nm in order to have a satisfactory thermal background.
- **Dark Current:** The measured dark current for 1.75 micron devices at 120K is in the neighborhood of 0.005 electrons/s/pixel. This is to be compared with roughly 0.03 electrons/s/pixel from the sky in the middle of the IR range. We would like to reduce this somewhat and will operate at 110K.
- **Flatness:** The chips are specified to have a flatness of 20 microns peak-to-valley, but in fact production chips are all flatter than 12 microns, and Teledyne is willing to accept this as a no-cost spec option.

- **Outputs:** 32 independent channels can be read at frequencies up to 400 kHz. Since there are 16 million pixels, there are 500,000 pixels per channel. At 100 kHz, then, a read takes 5 seconds. We will return to this later.

- **Read Noise:** The 1.75-micron material is inherently noisier than the more common 2.6-micron material. Teledyne have achieved 12 electron DCS (single-read) noise with the latter material, but the 1.75-micron material now yields noise of about 18 electrons for a single DCS read, and this is the figure we should plan on.

  The technique to reduce this in IR devices is *Fowler sampling*. Since the devices can be read non-destructively, one can read many times at the beginning and end of an exposure to establish charge levels. It is common to do 32 reads at each end, and obtain a "Fowler-32" result. This should result in a factor of 4 lower noise than a single DCS read, and the measured result of 3 electrons for the 2.6-micron devices is what one expects. For our devices, the expected number is 4.5 electrons, only slightly worse than the 4 electrons assumed in the strawman model. Another read technique is called up-the ramp, in which the chip is read continuously while the exposure is proceeding. This generates very large amounts of data but results in better noise performance than Fowler sampling (supposing one reads at the maximum rate in both cases) simply because one has more data, and in addition provides a mechanism for real-time rejection of cosmic ray events. We will investigate both options.

  For Fowler sampling, 32 reads at each end takes about 2.5 minutes at each end, which is acceptable, but we can probably do better. The read noise observed is very flat with read frequency, as shown in Figure 11, and we should be able to read with essentially no penalty at 200 kHz, and do Fowler-64 sampling, which should reduce our noise to just above 3 electrons.

- **Image Persistence:** Image persistence is a known problem with all HgCdTe devices. Teledyne has been working on this, and the newest devices are much better than the older ones. This is not really a significant problem with spectrographs in any case, but means that we *must* have a shutter, and must control the spectral samples so that excessively bright objects are not placed on the chip. The severity of the effect must be evaluated during test.

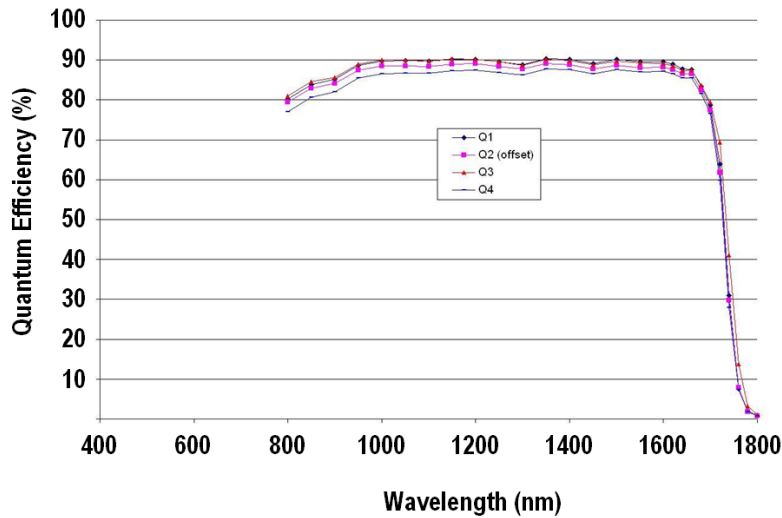

Figure 10  The quantum efficiency of three 1.75-micron Hg series detectors, using the same substrate-removed technology as the H4RG-15 uses.

The most significant issue with these detectors is actually acquiring them. They are very expensive (about $800K with the support electronics) and on-time delivery for small scientific customers has, understandably, not been a high priority for the company, given that they sell most of their devices to much larger and richer customers. They have promised devices on a 9-month ARO basis. We will do the best we can; there is no other source. We feel that it might be best *not* to push them for most of the devices, with the certainty that their processes will improve; our current plan is to get one

prototype device and order the 4 "flight" devices as late as possible. The prototype will serve as a spare (it will be full-cost, but we must have a spare).

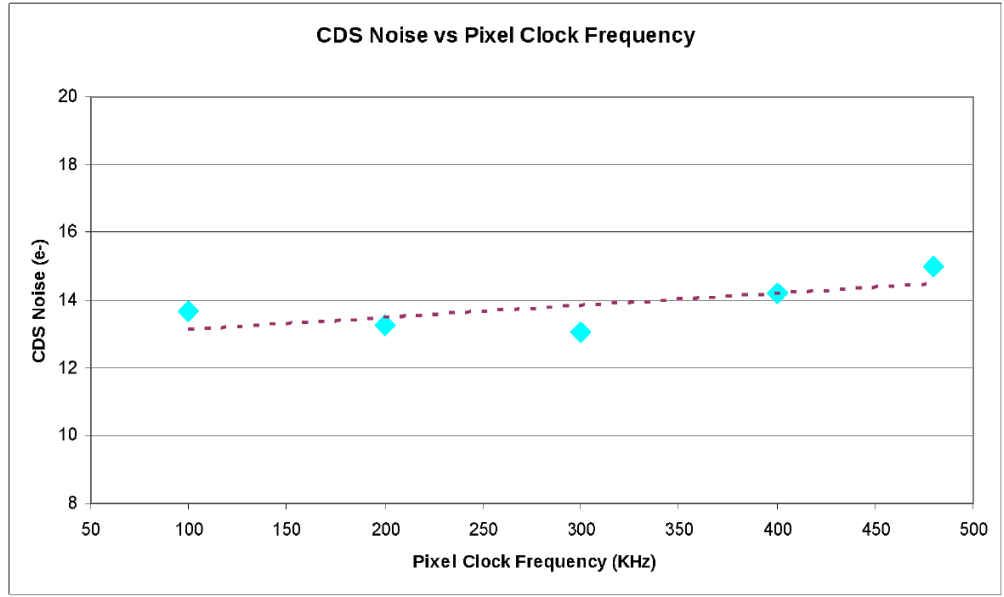

Figure 11  Single-read CDS read noise for a 2.6-micron H2RG device using a Sidecar signal processor versus pixel rate. Within the errors, there is no penalty for increasing the rate from the standard 100 kHz to 200 or even 300 kHz. The resulting data rates are larger, of course, because of the 32 channels: About a 10 MHz total pixel rate at 300 kHz. The data shown are for an engineering grade 2.5 um cutoff (SWIR) H2RG SCA; 77 K, 32 outputs, slow mode, buffered output, 250 mV bias with Sidecar ASIC focal plane electronics.

## 5.  THE THERMAL SYSTEM

### 5.1  General considerations for the thermal design

The thermal system for this instrument is somewhat unusual, because the cameras *are* the cryostats, and the approach to the system is of necessity somewhat different from the usual one of incorporating the detector in a relatively small dewar with a window to the outside world. One's first impression is that going to only 1.3 microns should not be sufficiently different from optical instruments that much care is necessary with thermal *background*, but that turns out not to be the case at all. In fact, the construction of the camera in the dewar is actually key to achieving the desired performance. It is to be noted that for this design, the whole camera assembly (mirror, detector, detector supports) is cold. In fact, the heat from the detector and radiation loads on the detector is carried away by the tripod support legs and removed by a cooler or coolers attached to the bases of those legs, which also cools the primary mirror. So the whole mirror/tripod assembly in this design is within a few degrees of the detector temperature. This is a very nice feature of the IR camera; it is not needed for the CCD cameras, but on the other hand is not a disadvantage. It is very clean, in that separate heat conductors for the detector, which would block light, are not required. Detector wiring can be run down the tripod legs.

It is a desideratum that the three cameras/dewars be as alike as possible to save manufacturing costs. Since it is the thermal considerations of the IR dewar which drives the design, we will discuss it first, and then show that we are not, in fact, driven to inordinate complication or expense by these factors - quite similar designs, mechanically almost identical, are possible for the CCD dewars. Though it might seem at first glance, since one does not need a cold environment for the CCDs, that the adoption of a common design is not optimal, but we have convinced ourselves that the economy of scale and the fact that we are on a very short schedule, in fact, makes this approach very desirable.

We first consider the thermal *signal* background for the IR detector. Crude calculations show that a detector, even with the relatively short 1.75 micron cutoff, has a *very* large background signal if it is looking at a room temperature (herein "room temperature" will mean 0 C, since that is an achievable lab temperature at Mauna Kea) source, even if at the focus of an f/1.1 camera which is completely cold outside the beam, about 40 electrons per pixel per second. This is to be contrasted with the signal from the sky continuum in PFS, which is in the neighborhood of 0.03 photons per pixel per second in the middle of the IR band. This clearly must be dealt with if we are to succeed. We clearly must live with whatever we get in the active wavelength range of the instrument unless we cool the spectrograph as a whole, which would be very, if not prohibitively, expensive. So the camera looks out at a room-temperature world. The exact view of the camera as it looks out into the spectrograph in the beam is quite complex; some solid angles and wavelengths go to the collimator and see back into the camera, but most terminate on some mechanical part of the spectrograph. In this discussion we will assume that all rays terminate on some 0C surface, either directly or by reflection from some other 0C surface. This will be a bit conservative, but actually not much; crude estimates place the overestimate of the background making this assumption in the 10 percent range. These same simple calculations show that for light shortward of the wavelength limit of the spectrograph, the background is about 3e-3 electrons/s, so if we could get rid of the radiation between 1.3 microns and the detector cutoff we would be fine; the background is only a tenth of the signal. So we need a filter, which we could coat onto the field flattener, immediately adjacent to the CCD or on one of the corrector surfaces. But if we make a filter which cuts at 1.3 microns, it will cut into the spectrograph range, since filters do not have sharp cuts, and, in addition, real filters do not block perfectly. If we specify a buildable filter, which has the characteristics

- A 50% cutoff wavelength of 1.32 microns,
- A 10% to 90% cutoff width of 400 Å,
- A rejection of 1.e-4 longward of 1.34 microns,
- A transmission of 95% in the 1-1.3 micron band

we get the results shown in Figure 12. In this figure *cumulative* fluxes are shown. At a given wavelength, the blue curves give the cumulative flux in electrons/s per pixel for all wavelengths *shorter* than that wavelength for two assumptions about the thermal rejection filter. The total background is given by the value at the far right. The red curve is the result of running the spectrograph at +10 C instead of 0 C; we *MUST* air-condition the spectrograph room or cool the spectrograph, but the former is almost certainly the desirable choice. One notices several things looking at these curves. First, the background is rising exponentially at the 1.3 micron cutoff wavelength, so the exact shape of the filter near the cutoff is very important. Second, the flux in the long-wavelength rejection part of the range is also becoming very large near the chip cutoff at 1.75 microns. It will be difficult to do anything about the first problem unless one does, in fact, cut into the desired spectral range a little bit, which one might like to do. Cutting at 1.28 microns instead of 1.30 lowers the background a factor of 2. However, with our cutoff, the background from wavelengths shorter than 1.5 microns is only 5.7e-3, a little more than half the total. The rest comes from leakage in the rejection region, again mostly at the very long-wavelength end. The dashed blue curve is the result for a rejection of 1.e-5, a background of about 6e-3 electrons/sec/pixel, which is probably attainable with a combination of filters - though the trapping of radiation between two dielectric filters will make this difficult.

Running the radiation shield at -80 C or colder results in negligible contribution from the shield even with no filtration on the field flattener, and we will probably run it colder; we neglect this source. Even without the better filters, we get a total background of 1e-2 electrons/s/pixel, a third of the sky signal. It seems very likely that this is a mild overestimate, and that we are likely to be able to achieve a background which is perhaps 20 percent of the sky level, around 7e-3.

But this does require a thermal shield - the detector must see cold surfaces at all angles which do not actually contain the beam - and those cold surfaces must extend all the way to the entrance aperture of the camera. The temperature of this shield must almost certainly be at -80 C (193 K) or below to produce acceptable backgrounds in the IR, and we will, as we will see below, run it *much* colder than this for other reasons. Once we have the shield designed, we will incorporate it into the CCD dewars as well, where it is not needed for background suppression, but does ease the thermal requirements somewhat.

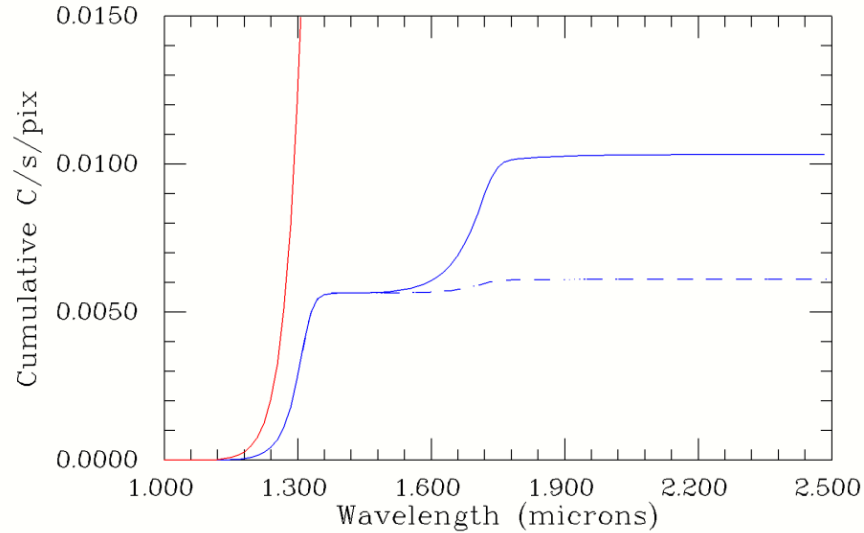

Figure 12 The IR background in electrons/s/pixel, plotted as the cumulative flux for wavelengths *shorter* than the indicated one. The curves are for the filter discussed in the text. The red curve is the background in the beam for a spectrograph temperature of +10 C, which is disastrously large. The blue curve is the same for 0 C, the design temperature. The dotted extension is for a reduced transmission in the rejection region, 1e-5 instead of 1e-4.

The detectors, again, must have temperatures which are basically determined by dark current considerations, and this demands CCD temperatures in the neighborhood of 160-170 K and the IR detector at 110-120 K. The latter, we will see, is a mild challenge.

**5.2 Cooling and total power**

Now that we have determined rough values for the temperatures of various parts of the dewar, we need to calculate heat fluxes and determine how much cooling power ("lift") we require to cool the system.

The main heat load is radiation from the corrector (unlike the signal background, the thermal load is mostly at wavelengths of ten microns or larger, at which the corrector is opaque and almost black), most of which ~ 90 percent) strikes the radiation shield. For the IR dewar, this is about 10 watts. Another ~ 4-5 watts will be lost through the insulation around the thermal shield and the back of the mirror, without which the thermal load is enormous (~ 400 W). So we know that we will need to pump about 15 watts from the dewar, but not all of this comes from very cold loads.

The detector and electronics receives radiation through the corrector, from the radiation shield, and generates some energy in its electronics. For the IR, the total is about 1.5 W.

This is carried away, again, by the tripod legs, in the current strawman design 0.5 cm by 2 cm, so with a cross-section of 1 $cm^2$, though only about half of that is copper. The primary heat load carried by the tripod, however, is radiation falling on the tripod itself from the corrector but mostly from the thermal shield. It must be very black in order to keep scattered light low, which almost certainly means that it will be very emissive at thermal wavelengths. This loss is mitigated by running the shield very cold, and the 'sweet spot' seems to be in the range 150-170 K; the detailed model we have constructed assumes a shield temperature of about 153 K, -120 C. At this shield temperature the radiation from the tripod to the shield is only a one watt, and the total load on the tripod is 1.1 W. For a detector temperature of 110 K, the mirror runs at 100 K, and the total load on the cooler which cools the mirror/detector is about 4.5 W. The thermal connection from the mirror to the cooler, accomplished with a rather massive copper or aluminum spider and silver strapping, drops another 5 degrees, so the cold head runs at 95 K.

A diagram showing the rough thermal configuration is shown in Figure 13.

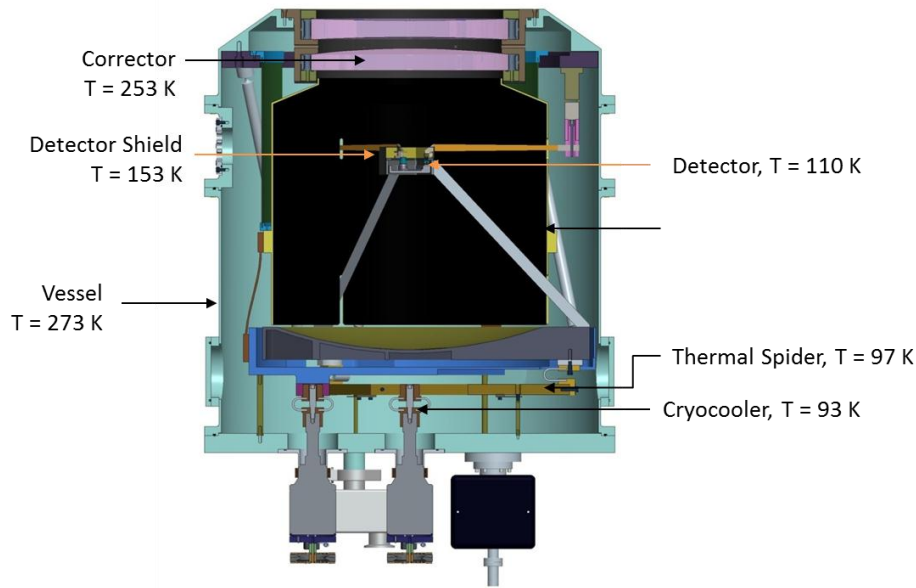

Figure 13 Cross-section of the IR cryostat showing the rough temperatures of the components. The detector mostly sees the radiation shield at -120 C (153 K). The whole mirror/detector assembly is cold, from about 110 K at the detector to about 100 K at the mirror to 95 K at the cold head of the main cooler.

The rest of the 15 W total comes from the radiation shield, and a much looser connection to ANOTHER cooler provides this. Since the shield runs at a much higher temperature than the detector/mirror assembly the higher load is actually easier on the cooler than the lower, colder load from the mirror/detector. This is not shown in detail in the drawing. For the CCD dewar, the requirements are much less stringent, and one cooler can easily handle the detector and the shield. The shield itself does not need to be very cold. The optimization does not have anything to do with background but just total power; the colder the shield the more power from the corrector it absorbs, but the less the legs of the tripod absorb from the shield. We have not done the optimization very carefully, but the temperature will be in the -20 C to -30 C range, and is not at all critical. The total load is only about 8 watts. We have pretty much settled on a cooler, a compact free-piston Stirling unit (the Cryotel GT) from Sunpower, which has 15 W of lift at 77 K. It is necessary to run two coolers on the IR camera, to have sufficient reserve - 15 W, even at temperatures a little higher than 77 K, seems much too close to the capacity of these coolers for comfort. As we have seen, the difficult cold load is only about 5 W, and the much higher shield load, about 10 W, is at a much higher temperature and is very easy on the cooler. The Sunpower cooler is *much* cheaper than the alternative linear pulse-tube coolers such as the Thales, which we have also investigated.

The relative advantages and disadvantages of the two cooler types are:

- The Sunpower is *much* cheaper; $12,000 including controller vs. $30,000 with no controller for the Thales.

- The Sunpower is *much* longer lived - at least 120,000 hours MTBF vs 20,000 hours for the Thales. If we used the Thales, we would have to have a scheme and resources to replace/repair them at least every 18 months; the Sunpower units will last 10 years. A Sunpower unit of this type has been operating without difficulty in a spacecraft for 10 years on orbit.

- The Sunpower unit uses somewhat less power; 240 W vs. 350 W for the Thales.

- The Sunpower units vibrate quite badly; the Thales does not. Various vibration mitigation schemes exist, and we will explore them. If the vibration problem can be solved, it appears to be the best solution in every way. The vibration with their standard tuned mechanical damper is about 0.4 G at 60 Hz for the unit hanging free, which corresponds to an amplitude of about 28 microns. Dynamic electrical damping can reduce this to about

40 mG, 3 microns, which is probably OK without any complex mechanical vibration absorption, but requires some work. The dynamic transducer is available, but the controller is still under development by Sunpower.

## 6. THE VACUUM SYSTEM

The vacuum system required for PFS is relatively straightforward, the only complication being the large number of dewars (12), their relatively large volume (240 L), and the operational requirements of bringing the whole system down and back up relatively quickly. We must reach full vacuum so the detectors can be run in not more than a few days, and the vacuum level must be of order one microtorr. The lifetime of ion pumps, which we will use to maintain the vacuum, depends on the ultimate vacuum, and is several years (80,000 hours) at one microtorr. The thermal performance also degrades significantly at poorer vacuum levels, so one microtorr long term is a reasonable goal.

The contents of the dewar are mostly bakeable, the detectors and electronics being the only subsystems sensitive to high temperatures, and the electronic systems are small. Compared to HSC, for example, with electronics capable of supporting more than a hundred CCDs all the way to digital output and all in the vacuum, the system is very clean.

The current plan is to have four vacuum ports on the rear face of the cryostat, two of which will have conflat flanges and be permanently equipped with Varian VacIon 20 star-cell 20 L/s ion pumps, another with a valve to which can be mated the roughing and pump-down pumps. The roughing will be done with portable scroll pumps and the pumpdown with a ~ 80 L/s turbomolecular pump. These pumps are standard items and we have not yet chosen which ones we will use. The fourth will mount permanently a standard combination gauge, probably the Pfeiffer one, a very robust unit which we have used in the past.

Whether we will have to have as many turbopumps as dewars is not yet clear, and will depend on the average pumping time-that is, how long a dewar must remain on the turbopump before the ion pumps can be turned on safely. If this is sufficiently short, we can swap turbos among the dewars, but if it is long, the prudent thing to do would be for each dewar to have a dedicated turbopump. The cost for a turbopump and controller is in the neighborhood of $7000, for a scroll pump for the roughing about $10,000. The ion pumps and their controllers cost in the neighborhood of $5000 each, so the whole vacuum system costs roughly $250,000 plus labor.

## 7. THE ELECTRONICS

### 7.1 The in-dewar IR electronics

The only element needed in the dewar for the IR chips is the `"Sidecar" module on a small circuit card, the only other components (also on the card) being decoupling capacitors for the sidecar. This unit is shown in Figure 9, not mounted on its board. This package and the board is not yet commercially available, but is promised in the first quarter of 2012. It will use a relatively new technology called pin-grid-array, PGA, which is similar to the standard ball-grid-array (BGA) mounting which is used for microprocessors and other products with very large numbers of contacts in a dense array, but the balls are replaced with relatively thin columns; the result is a system with greatly improved tolerance to the thermal mismatch between the ceramic package and the composite laminate circuit board. There is a commercially available product which does not use a sealed package and is on a considerably larger board (but which is still smaller than the detector) which we could use if necessary.

The sidecar will be mounted immediately adjacent to the detector, either directly on the detector connector, as shown in the previous figures, or on a very short flexible printed circuit. The Sidecar is relatively happy at almost any operating temperature, and we will have to do more detailed thermal modeling than we have done to determine the operating temperature for it that we can easily reach. It dissipates only about 150 mW at 100 kHz readout rate; that power scales very nearly linearly with the frequency. The wiring to the outside world from the sidecar will go down one of the tripod legs and then to the connectors on the front ring. It requires only 3.3 VDC at very small current to operate, and generates complete telemetry of voltages and operating state internally.

## 7.2 The in-dewar CCD electronics

The CCD electronics will be patterned after the SDSS CCD electronics,[7] which has been used successfully for 15 years at APO, but will be redesigned with more modern components and mounted more compactly. The electronics in the dewar will be divided into two boards or board sets, one which is mounted immediately adjacent to the CCD and simply serves as a preamplifier and impedance transformer for the CCD signal. It uses one fast, low-noise AD1810 operational amplifier per channel output from the CCD, and a very low-leakage, low-injection switch, the ADG1236-1, for DC restoration. These are both obtainable in very small packages, and these and their necessary bypass and coupling capacitors will fit on one or at most two roughly 50 mm square boards which will be mounted directly behind the CCD. Coming from this preamp will be 8 video lines, and serving it will be power and a CMOS signal to run the switches, which are used for DC restoration. A photo of the similar board used in SDSS-1 is shown in Figure 14; it handled only two video channels, but had all the clock driver functions as well.

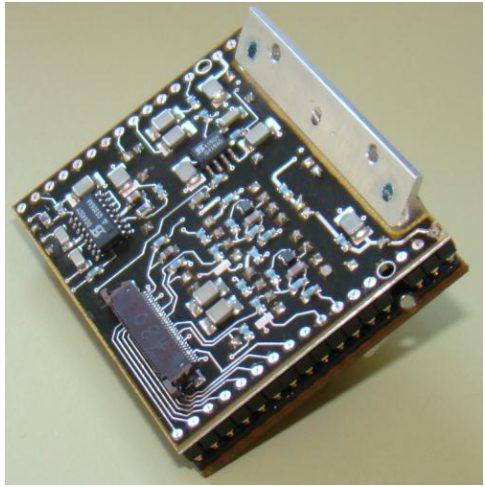

Figure 14  Photograph of the in-dewar preamplifier module used in SDSS-I in the imager and spectrographs, It is a double-board stack 40x50 mm and handles two channels of video and the clock drivers for all the CCD clocks.

Serving this preamp board and the CCD itself will be three modules mounted to the front ring on posts which extend back to the vicinity of the base of the tripod. One of these does the double-correlated sampling and digitization. Another generates the voltages necessary for the operation of the CCD, and the third generates the clock waveforms. The wiring to them (in the form of low-thermal-conductivity flexible printed circuits) to them will run up the corresponding tripod leg to the detector.

The output from these modules are 8 serial lines of digital data and an analog voltage bus for monitoring, and the input from outside is a set of LVDS clock and control signals and power. These are routed to the hermetic connectors on the front ring and thence to the external electronics.

## 7.3 Interfacing the sidecar-electronics and software

Teledyne currently sells an interface board which communicates via USB to a Windows host and provides software to acquire and store data. They are developing and will market (promised again in 1st quarter 2012), a new board which is similar but is controlled and moves data over gigabit ethernet. We discuss the computer which talks to the interface card below; the preliminary notion is that one machine will handle all the cameras on one spectrograph, and, hopefully, all the housekeeping tasks for the spectrograph.

### 7.4 Interfacing the CCD electronics and software

This task is much more involved than that for the IR chip. The SDSS control system is so old and uses so many obsolete parts that it must be redone. On the one hand, the physical limitations that made doing the SDSS system hard are now not relevant at all, but on the other hand, there is a fair amount of software/firmware which needs to be implemented.

It is also clear that we can reuse a fair amount of the technology developed for HSC, and it is quite possible that the HSC back-end (BEE) modules can be used absolutely as-is. The in-dewar front-end (FEE) boards are too large (they are designed for running many CCDs simultaneously), but the design is very nice and can certainly be adapted without difficulty.

### 7.5 Data system, data storage, and Subaru interface

Both kinds of camera controllers will output data on gigabit ethernet, and will be controllable over that same link. These three data lines will go to a Linux control computer with at least 4 gigabit ethernet ports, three of which are used for the three detector systems and the fourth for communication with Subaru.

A fifth port, which could be quite slow, will be used for housekeeping tasks, including the shutter, back illumination system, instrument temperature monitoring, cooler control, vacuum and perhaps vibration monitoring, etc.

The data rates are quite low by today's standards. The IR detectors are read while exposing, and the rate will not exceed 300 kpixels/sec for 32 channels, or about 10 MBytes/sec, 100 Mbits/sec. The 16 channels of the two CCD cameras read at 75 kpixels/sec produce a data rate in a 27 second burst of 1.2 MBytes/sec, 12 Mbits/sec, and the optical and IR data are not produced simultaneously.

The use of standard ethernet protocols will allow us to program the interface at relatively high level in C/C++ for a Linux computer. The initial plan is to stage all the data on this machine on either spinning or solid-state disks. If we do up-the-ramp for the IR detector at 100 kpixels/sec, we generate about 200 GB of data in an 8-hour night; if at 300 kpixels/sec, three times that, 600 GB. The optical data are small compared to this, with all calibration data probably 10 frames per hour, about 5 GB per night. It is *probably* not necessary to archive all the up-the-ramp data, but the cost of doing so is not prohibitive.

### 7.6 The utility system

The housekeeping functions will depend partly on an interface with LAM, who will probably either provide a microcontroller which talks to the spectrograph utilities or provides digital/analog signals to us for combination with our utility signals. It would make sense, however, for this system to be closely integrated with the control system discussed in the last section, so that one machine can run the spectrograph as a unit, from control and monitoring to data acquisition. The functions of this system include:

- Monitor all spectrograph temperatures, including passive temperatures on the spectrograph body and optics mountings and temperatures of the active devices, the coolers and ion pumps.

- Monitor the flow of the cooling fluid for the coolers.

- Monitor the flow of the dry air or nitrogen for the spectrograph purge.

- Monitor all ion pump currents for vacuum, and monitor the separate vacuum gauges if any.

- Monitor the vibration level on the coolers.

- Control the shutter and monitor its state.

- Control the back-illumination system, interlock it with the shutter, and monitor its state.

## 8. SUMMARY

We have described here the detectors and dewars for the PFS spectrograph, consisting of three almost identical vacuum Schmidt cameras for the blue, red, and infrared arms of each of the spectrographs. We have shown that the optical and thermal requirements are manageable without significant engineering difficulty, and no thermal or electrical consideration causes significant performance degradation beyond what the natural backgrounds and irreducible read noise in the detectors impose a priori. These are, in turn, consistent with the science requirements, if we blindly pretend that the requirements flow that direction instead of the truth, which we all know, that they flow the other way.

## ACKNOWLEDGEMENTS

We gratefully acknowledge support from the Funding Program for World-Leading Innovative R&D on Science and Technology (FIRST) "Subaru Measurements of Images and Redshifts (SuMIRe)", CSTP, Japan.